\documentclass{iopart}
\input epsf
\begin{document}
\jl{1}
\title[Mixed Potts ferro-antiferromagnets]
{Kosterlitz-Thouless transition in three-state mixed Potts 
ferro-antiferromagnets}
\author{Miguel Quartin and S L A de Queiroz\footnote{Corresponding 
author}}

\address{Instituto de F\'\i sica, UFRJ, Caixa Postal 68528, 
21941--972 Rio de Janeiro RJ, Brazil}

\begin{abstract}
We study three-state Potts spins on a square lattice, in which all 
bonds are ferromagnetic along one of the lattice directions, and
antiferromagnetic along the other. Numerical transfer-matrix are 
used, on infinite strips of width $L$ sites, $4 \leq L \leq 14$.
Based on the analysis of the ratio of scaled mass gaps (inverse 
correlation lengths) and scaled 
domain-wall free energies, we provide strong evidence that a critical
(Kosterlitz-Thouless) phase is present, whose upper limit is, in 
our best estimate, $T_c=0.29 \pm 0.01$. From analysis of the (extremely 
anisotropic) nature of excitations below $T_c$, we argue that the
critical phase extends all the way down to $T=0$.
While domain walls parallel to the ferromagnetic
direction are soft for the whole extent of the critical phase, those along
the antiferromagnetic direction seem to undergo a softening transition at
a finite temperature. Assuming a bulk correlation length varying, for
$T>T_c$, as
$\xi (T) =a_\xi\, \exp \left[ b_\xi\,(T-T_c)^{-\sigma}\right]$, $\sigma
\simeq 1/2$, we attempt finite-size scaling plots of our finite-width
correlation lengths. Our best results are for $T_c=0.50 \pm 0.01$. We 
propose a scenario in which such  inconsistency is attributed to the 
extreme narrowness of the critical region.
\end{abstract}

\pacs{05.20.-y, 05.50.+q, 64.60.Fr, 75.10.Hk}


 
\section{Introduction} 

Many spin systems exhibit exotic low-temperature
phases, frequently as a result of competing interactions coupled with
suitable ground-state degeneracy. In this paper we study a borderline case
in which, although neither frustration nor macroscopic
ground-state entropy are present, a critical phase (i.e. with
power-law decay of correlations against distance) arises for a finite 
temperature range above zero. The prototypical system in which such sort 
of phase
occurs is the two-dimensional $XY$ model~\cite{kt}.
The corresponding transition from the high-temperature (paramagnetic)
to the critical phase, characterized by an exponential divergence of
the correlation length, is known as  Kosterlitz-Thouless (KT) transition.
Although the continuous symmetry of $XY$ spins plays an important role
in the formation of vortices, whose presence is an essential feature 
of the low-temperature behaviour, it was soon realized
that critical phases could be sustained also in some 
two-dimensional magnets with discrete spin symmetry. Examples are $Z(N)$ 
models with $N>4$~\cite{jose,eps}, and models with a 
macroscopically degenerate ground state to which suitable 
degeneracy-lifting interactions are added. The latter case is 
illustrated by the triangular Ising
antiferromagnet in a uniform magnetic field~\cite{nhb,bnw,bn93}, or
with ferromagnetic second-neighbour interactions~\cite{dswb,lan,dqd95}.

We consider the uniformly anisotropic Potts model with nearest-neighbour 
couplings on a square lattice~\cite{wu}, whose Hamiltonian reads:
\begin{equation}
{\cal H} = - \sum_{x,y} \left(\,J_x\,\delta(\sigma_{x,y}-\sigma_{x+1,y})+
J_y\,\delta(\sigma_{x,y}-\sigma_{x,y+1})\,\right)
\label{eq:1}
\end{equation}
where the Potts spins $\sigma_{x,y}$ take any one of $q$ values, and
the lattice site coordinates are given by $(x,y)$.

Our interest is in the mixed (ferro-antiferromagnetic) 
case~\cite{wu,ost,ksw} $J_x >0$, $J_y < 0$. For any $q$, the Hamiltonian 
is unfrustrated, as the product of interactions around plaquettes is 
always positive.
The ground state of an $L \times L$ system has degeneracy ${\cal 
O}\,(q-1)^L$; thus, even though this diverges in the thermodynamic limit 
for $q >2$, the  residual entropy per spin approaches zero as  $L^{-1}$.

The $q=3$ mixed model which will concern us here has attracted
special attention~\cite{ost,ksw,hm83,ttt,sw87,yasu87,fg02}, as 
it is believed to exhibit a KT transition.
Furthermore, evidence has been found  that no transition occurs for 
$q>3$~\cite{fg02}. These two findings may in fact be viewed as 
complementary~\cite{fg02}. In previous work, a great deal of effort
was dedicated to calculating the critical temperature $T_c$ above which
the system is paramagnetic, particularly the variation of $T_c$ against 
the ratio $J_x/J_y$, with a rather large degree of scatter in the 
corresponding results~\cite{ksw,ttt,sw87,yasu87,fg02}. Although, 
in these latter references, assorted evidence is quoted in favour of the
view that the transition is indeed ``of an unconventional 
type''~\cite{ksw}, a more precise characterization is attempted only in
~\cite{sw87}, where the correlation lengths from Monte Carlo studies of $L 
\times L$ systems, $L \leq 60$, are shown to fit reasonably well    
to the exponential form in $(T-T_c)^{-1/2}$, expected to hold close to a 
KT transition~\cite{kt}. An investigation of the problem in its
$(1+1)$-dimensional quantum version~\cite{hm83} lends numerical
support to the respective form for the case, namely the mass gap
vanishing as $\exp\,(-a\,(\lambda-\lambda_c)^{-\sigma})$, where
$\lambda$ is the temperature-like coupling. However, the combined effects
of small system sizes and uncertainty in the determination of $\lambda_c$
allow the authors of ~\cite{hm83} to conclude only that $\sigma < 0.9$.   

As regards the low-temperature region, the critical (or massless) phase
is analysed in some detail in~\cite{hm83}; upon consideration of the
Roomany-Wyld approximants~\cite{rw} to the beta-function~\cite{ns82,dds}, 
it is  deemed as plausible that the mass gap vanishes all the way from
$\lambda=0$ to $\lambda_c$ (corresponding, in the classical analog,
to a critical phase from $T=0$ to $T_c$). In a Monte Carlo 
renormalization group study of the chiral Potts model~\cite{hjb}
(with the parameters set in such a way as to map onto the present 
model~\cite{ost}), the correlation functions are shown to behave 
consistently with the existence of a critical phase. However, the 
temperature range considered in~\cite{hjb} goes only halfway down to zero
from $T_c$. 

Thus, it seems worthwhile investigating the behaviour of this system in 
more detail, both (a) at the transition and (b) at low temperatures.
In connection with (b), it must be recalled
that, for triangular Ising antiferromagnets with second-neighbour
ferromagnetic interactions~\cite{dswb,lan,dqd95}, the critical phase
does not extend all the way down to $T=0$: there is an ordered phase
with nonzero magnetization for a finite temperature range {\em below}
the KT phase. While strictly similar behaviour should not be 
expected in the present case (since here the ground state has zero
macroscopic magnetization), the question still remains of precisely how
the (highly anisotropic) correlations spread at low temperatures. In
~\cite{ost,ksw}, arguments are given to show that slightly above $T=0$
an effective ferromagnetic coupling arises between second-neighbour 
rows of ferromagnetically coupled spins. This would 
single out a small subset of row configurations, namely $010101$, $020202$ 
and $121212$~\cite{ost} as energetically more favourable. It appears 
that the effect of this on the free energy (hence, on the 
corresponding thermodynamic state) has not been quantitatively 
investigated so far.  

We apply numerical transfer-matrix (TM) methods to the $q=3$ mixed Potts
model of equation~(\ref{eq:1}), on strips of a square lattice of width
$4 \leq L \leq 14$ sites with periodic boundary conditions  across
(except where domain-wall free energies are analysed, in which case
both periodic and twisted boundary conditions are considered).
We shall restrict ourselves to $J_x/J_y=-1$.

In Section~\ref{sec:smg} we consider the scaling of correlation lengths,
as obtained from the two largest eigenvalues of the TM. along both
directions $x$ and $y$ of equation~(\ref{eq:1}). From these we
calculate the ratios of scaled gaps, 
which provide relevant information on the
extent of the critical phase. In  Section~\ref{sec:dws}, domain-wall
free energies are calculated, and their associated correlation lengths
are investigated; in Section~\ref{sec:tc} the region near the
transition from  critical to paramagnetic behaviour is examined, 
especially as regards the form in which the thermodynamic correlation
length diverges as $T \to T_c^+$; finally, in  Section~\ref{sec:conc},
concluding remarks are made.

\section{Correlation lengths}
\label{sec:smg}
Here we consider the largest correlation length $\xi_L(T)$ on a strip 
of width $L$, {\it i.e.} the one given by $\xi_L^{-1}(T)= \ln 
(\Lambda_0/|\,\Lambda_1\,|)$ where $\Lambda_{0,1}$ are the two largest 
eigenvalues of the  TM in absolute value. Thus it picks up the slowest 
decay of correlations along the specified direction of iteration of the TM. 

With the correlation function $G(r) \equiv \langle \delta (\sigma_0 - 
\sigma_r)\rangle -1/3$~\cite{sw87}, for the ferromagnetic direction
one has $G(r) \sim \exp  (-r/\xi^F)$, while in the antiferromagnetic 
direction $ G(r)\sim (-1)^r\,\exp (-r/\xi^{AF})$.

As remarked in \cite{fg02}, the phenomenogical-renormalisation equation
for the critical temperature,
\begin{equation}
L\,\xi_L^{-1}(T^\ast)=L^\prime\,\xi_{L^\prime}^{-1}(T^\ast)\ ,
\label{eq:2}
\end{equation}
only admits a fixed point $T^\ast$ when the TM is taken along the 
ferromagnetic direction. 
\begin{figure}
\epsfxsize=8.5cm
\begin{center}
\epsffile{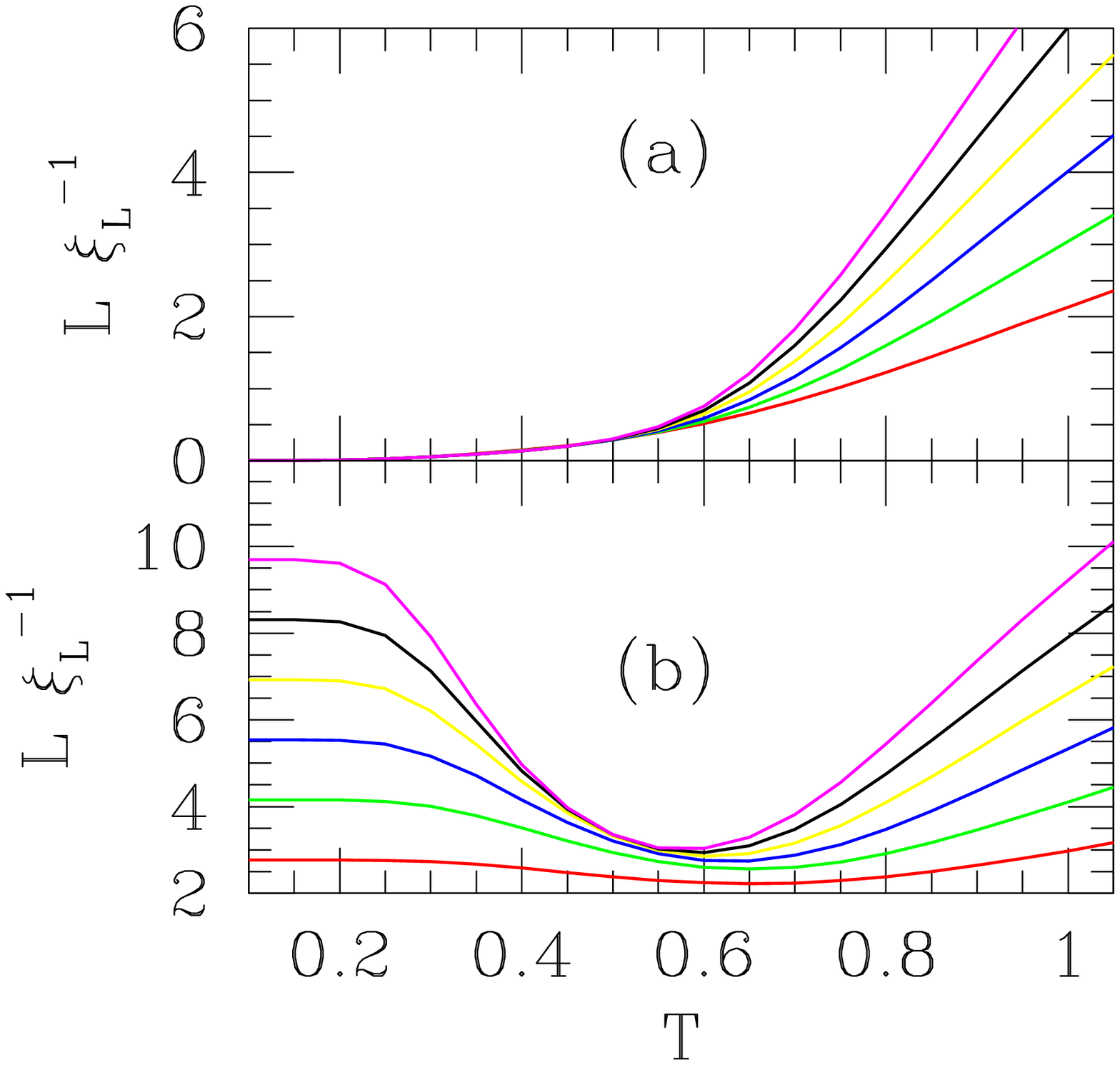}
\caption[]{
$L\,\xi_L^{-1}$ against temperature: (a) TM in the ferromagnetic 
direction; (b) TM in the antiferromagnetic direction. Bottom to top:
$L=4,6, \dots 14$.
}
\label{fig:1}
\end{center}
\end{figure}
The raw data for $L\,\xi_L^{-1}$ against temperature
are shown in Fig.~\ref{fig:1}, respectively (a) for the TM taken in
the ferro-- (F) and (b) antiferromagnetic (AF) directions. In the latter
case, curves for consecutive strip widths get almost tangent as $L$
grows, in the region  $T \sim 0.5-0.6$ (the $L=12$ and $14$ curves reach 
within one part in $10^3$ of each other, at $T=0.5$)  . A  clearer 
understanding of both sets of data can be gained as follows.

\begin{figure}
\epsfxsize=8.5cm
\begin{center}
\epsffile{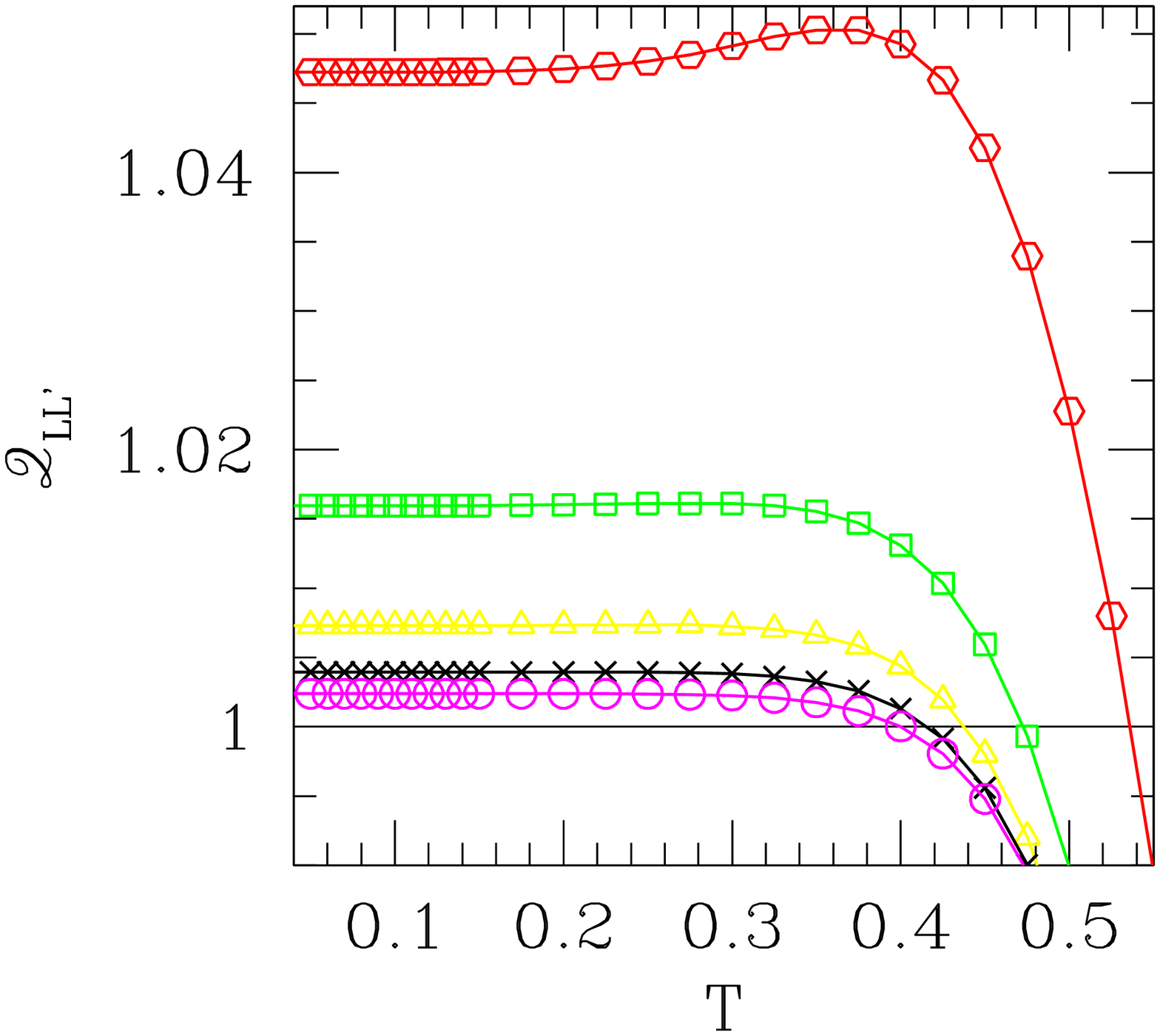}
\caption[]{${\cal Q}_{L\,L^\prime}(T)$ of equation~\protect{(\ref{eq:3})}
against $T$, for $L^\prime=L+2$. Top to bottom: $L=4, 6,\dots 12$ 
}
\label{fig:2}
\end{center}
\end{figure}

For the case of Fig.~\ref{fig:1} (a) we plot the ratio of scaled mass 
gaps~\cite{dqd95,rw,hb81,sd83,ds84}
\begin{equation}
{\cal Q}_{L\,L^\prime}(T) \equiv~\left[\,L\,\xi_L^{-1}(T)\,\right]/
\left[\,L^\prime\,\xi_{L^\prime}^{-1}(T)\,\right]\ ,\qquad {\rm (F)}  
\label{eq:3}
\end{equation}
where we always take $L^\prime=L+2$. At the fixed point of 
equation~(\ref{eq:2}),
${\cal Q}_{L\,L^\prime}(T^\ast)=1$, and one expects it to remain close to 
this value for an extended temperature interval if a critical phase is 
present~\cite{dqd95,rw,hb81,sd83,ds84}. 
Indeed this is what happens here, as shown in Fig.~\ref{fig:2}. Starting
slightly below the respective $T^\ast$, and down to the lowest
temperatures reached before running into numerical 
problems (typically $T=0.03-0.04$), the curves remain rather flat. The 
values of ${\cal Q}_{L\,L^\prime}$ at which they stabilize are, 
respectively, $1.04724$, $1.01596$, $1.00730$, $1.00394$, $1.00239$, for 
$L=4, 6, \dots 12$. This sequence behaves as ${\cal 
Q}_{L\,L^\prime}=a+b\,L^{-x}$,
where $x \simeq 2.7$ and $a$ equals unity within one part in $10^3$.
As regards the extent of the flat region in the thermodynamic limit, the
shape of curves in Fig.~\ref{fig:2} suggests that the upper limit $T_c$ be
given by extrapolation of the sequence of fixed points of
equation~(\ref{eq:2}): $T_c =\lim_{L\to \infty} T^\ast_{L\,L^\prime}$. 
Such
sequence is the same given in Table 1 of~\cite{fg02}, with the additional
point $T^\ast_{12,14}=0.40030$. Using $L=4, 6,\dots 12$ we get
$T^\ast_{L\,L^\prime} =T_c+c\,L^{-y}$ with $T_c=0.29 \pm 0.01$ (to be
compared with $T_c=0.32 \pm 0.03$~\cite{fg02}), $y \simeq 0.7$.
Thus we have convincing evidence that the system exhibits a critical
phase, which extends down from $T_c \simeq 0.3$, at least to the lowest
temperatures examined, and presumably all the way down to $T=0$.

\begin{figure}
\epsfxsize=8.5cm
\begin{center}
\epsffile{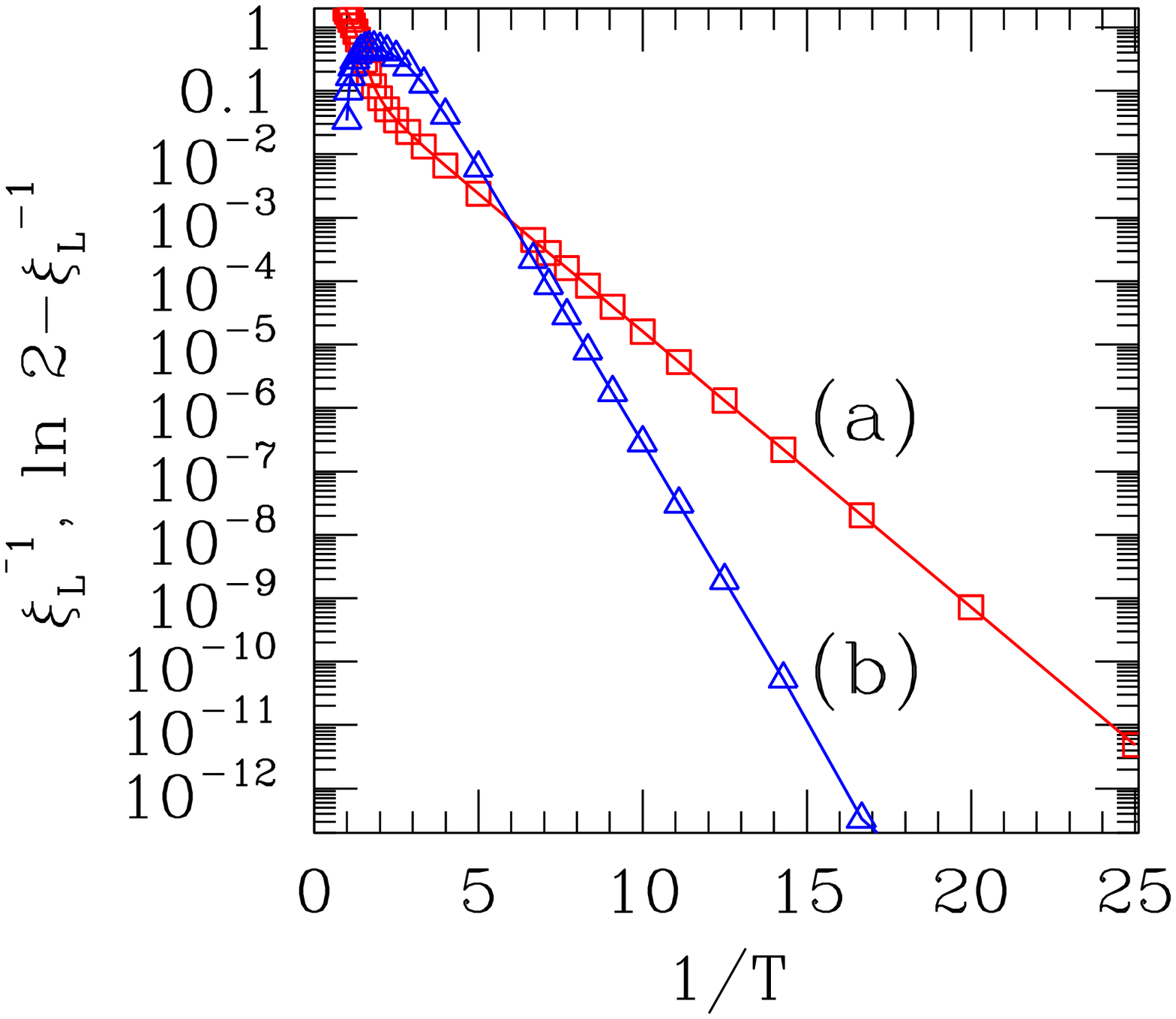}
\caption[]{Curve (a): $\xi_L^{-1}$ against $1/T$, $L=14$, TM in the 
F direction. Squares are calculated points; connecting curve approaches
behaviour $\sim \exp(-1/T)$ from $1/T \simeq 3$ onwards.
Curve (b): $\ln 2 -\xi_L^{-1}$ against $1/T$, $L=14$, TM in the
AF direction. Triangles are calculated points; connecting curve approaches 
behaviour $\sim \exp(-2/T)$ from $1/T \simeq 3$ onwards.  
}
\label{fig:3}
\end{center}
\end{figure}

In order to check on this latter assumption, it is instructive to
look at the low-temperature behaviour of the correlation lengths 
themselves. This is illustrated in curve (a) of Fig.~\ref{fig:3} which 
shows that, immediately below $T_c$ the functional 
form $\xi^{-1} =F(L)\,\exp(-1/T)$ (with $F(L) \sim L^{-1}$, from the
results for ${\cal Q}_{L\,L^\prime}(T)$ above) sets in 
along the F direction. This 
corresponds to the lowest-energy ($T \sim 0$) excitations in that 
direction, which are obtained by breaking a single ferromagnetic bond, eg 
$\dots 000000 \dots \to \dots 000111 \dots$ (while, in this example,
both neighbouring chains are $\dots 222222 \dots$). See 
below  the  contrasting case for the AF direction. In other 
words, $T \to 0$ behaviour is dominant all the way down from $T_c$, 
consistent with the idea that there is a single (critical) phase from
$T=0$ upwards to $T \simeq 0.3$.

Examination of data taken with the TM along the AF direction, 
Fig.~\ref{fig:1} (b), is helped because the zero-temperature bulk
correlation length in that direction is known exactly~\cite{ksw}:
\begin{equation}
\xi^{-1} (T=0) = \ln 2\ .\qquad {\rm (AF)}
\label{eq:4} 
\end{equation}
We then look at the finite-temperature and finite-size deviations
of $\xi^{-1}_L(T)$ from equation~(\ref{eq:4}), as displayed in curve (b) 
of Fig.~\ref{fig:3}. 
We have been able to reach roughly the same temperatures (usually $T 
\sim 0.03$) as with the TM along the F direction, 
before running into numerical problems. The trend shown for $L=14$
in the Figure is followed for all strip  widths, namely below $T \simeq 
0.3$ one has 
\begin{equation}
\ln 2 - \xi^{-1}_L (T)  = a(L)\,\exp (-2/T)\ .\qquad {\rm (AF)} 
\label{eq:5}
\end{equation}
We have found empirically that the $L-$dependent prefactor is well
fitted by a form $a(L)=a_0\,\exp(-b/L)$, $a_0 \simeq 500$, $b \simeq 
19.3$, except for $L=4$ which is off by $\sim 50\%$. 

The factor of two in the exponential reflects the single-spin nature 
of elementary excitations along the AF direction, as the least costly move 
in this case consists in overturning a spin (breaking both its 
ferromagnetic bonds) 
without breaking any of its antiferromagnetic ones. Again, the fact that 
such low--$T$ behaviour takes over immediately below $T_c$
indicates that there is a single phase between $T=0$ and $T_c$.
 
\section{Domain-wall scaling}
\label{sec:dws}
For bulk $d-$dimensional systems, hyperscaling implies that close to the 
transition, the domain-wall  free energy $\Delta F$
vanishes as the correlation length $\xi$ diverges, 
according to $\Delta F \sim \xi^{-(d-1)}$~\cite{watson}.
Assuming power-law singularities, $\xi \sim 
|T_c-T|^{-\nu}$, $\Delta F \sim(T_c-T)^\mu$, $T \to T_c^-$, one has
$\mu=(d-1)\nu$~\cite{watson}.
Specializing to $d=2$ and using standard finite-size 
scaling~\cite{barber}, one can derive the
basic fixed-point equation of domain-wall renormalisation 
group on strips~\cite{mcmillan}:  
\begin{equation}
L\,\Delta f_L(T^\ast)=L^\prime\,\Delta f_{L^\prime}(T^\ast)\ ,
\label{eq:6}
\end{equation}
where $\Delta f_L$ is the free energy per unit length, in units of $T$,
of a seam along the full length of the strip: $\Delta f_L = 
f^T_L-f^P_L$, with
$f^P_L$ ($f^T_L$) being the corresponding free energy  for a strip
with periodic (twisted) boundary conditions across. 

While, for Ising spins, twisted and antiferromagnetic boundary conditions 
coincide~\cite{mcmillan}, careful consideration~\cite{pdN88}
shows that in Potts systems with $q > 2$ this does not hold. 
Indeed, in order to produce a seam along the strip 
one needs an interaction which {\it favours} pairing between well-defined
(and different) spin states, {\it e.g.} $1-2$, $2-3$, $3-1$. An 
antiferromagnetic Potts interaction only makes selected pairings
energetically {\it unfavourable}, while leaving all other combinations
equally probable.

\begin{figure}
\epsfxsize=8.5cm
\begin{center}
\epsffile{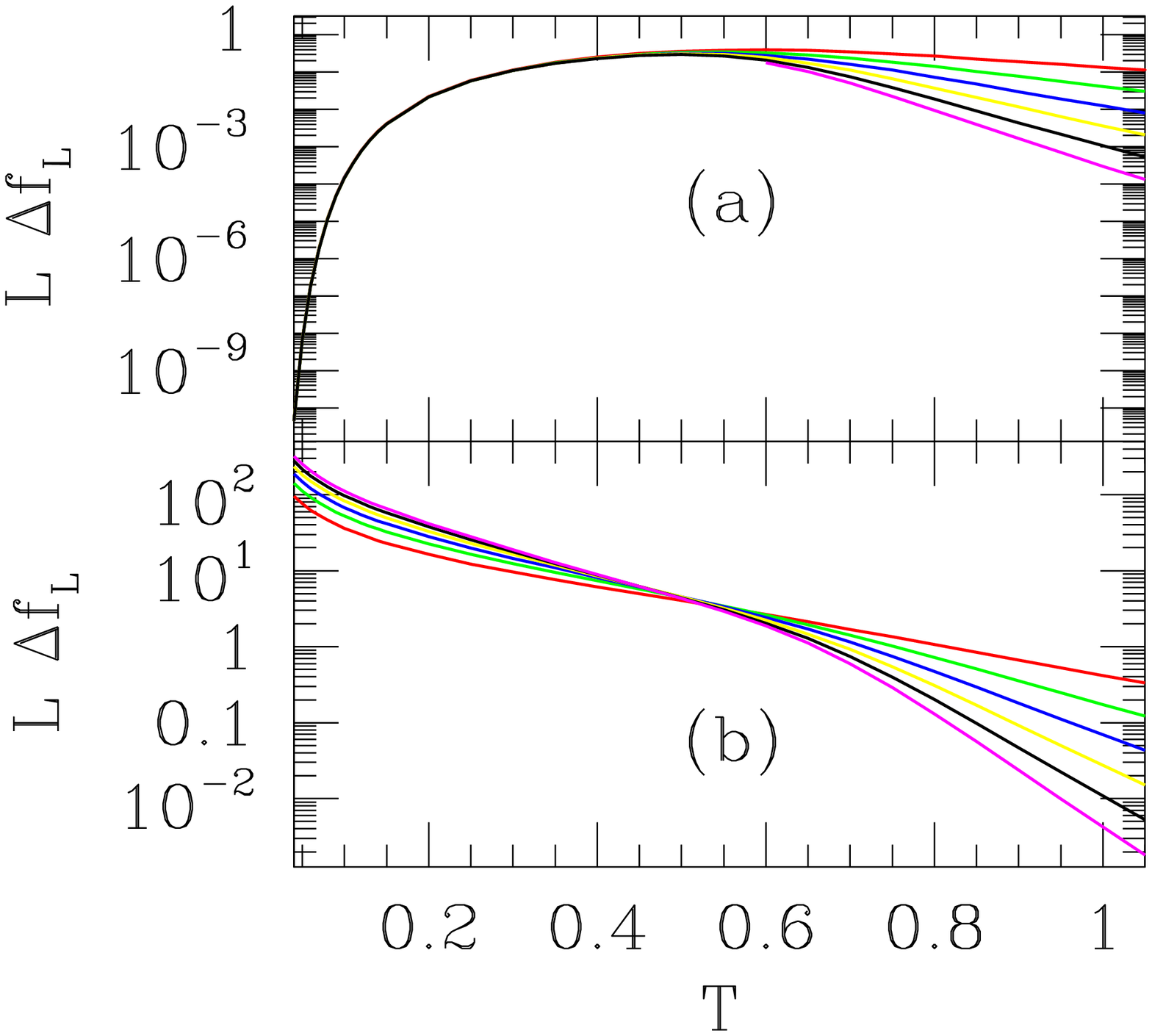}
\caption[]{
$L\,\Delta f_L$ against temperature. (a): TM in the ferromagnetic 
direction. On this scale, curves are indistinguishable up to $T \simeq 
0.4$; top to bottom (on right-hand side): $L=4,6, \dots 14$. (b) TM in 
the antiferromagnetic direction. Curves cross in the 
central region of $T$ axis. Top to bottom (on right-hand side):
$L=4,6, \dots 14$.
}
\label{fig:4}
\end{center}
\end{figure}

Therefore, for Potts models
with $q>2$, twisted boundary conditions are obtained by changing
the interaction across the $L-$ direction as follows~\cite{pdN88}:
\begin{equation}
J\,\delta(\sigma_{1}-\sigma_{L})\ \to 
J\,\delta(\sigma_{1}-\sigma_{L}+1\,({\rm mod}\ q))\ .
\label{eq:7}
\end{equation} 
With the above definitions, one has $\Delta f_L = -\ln (\Lambda_0^T /
\Lambda_0^P)$ where $\Lambda_0^T$, $\Lambda_0^P$ are the largest 
eigenvalues of the TM, respectively with periodic and twisted boundary 
conditions across.
In the present case, different results are to be expected, depending on 
whether the TM is set along the F or AF directions.  In 
Figure~\ref{fig:4}
we show $L\,\Delta f_L$ against temperature for both cases.

\begin{figure}
\epsfxsize=8.5cm
\begin{center}
\epsffile{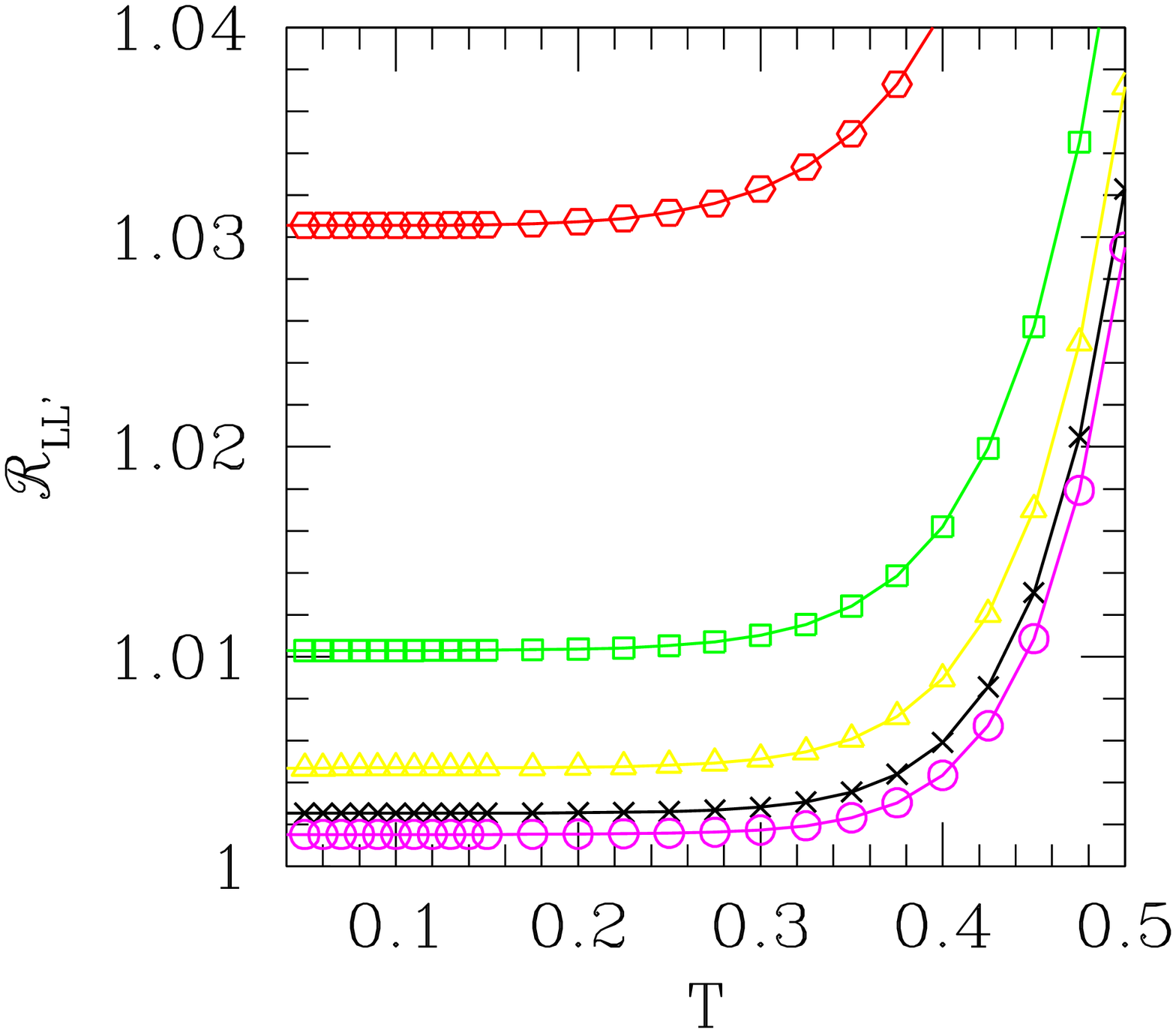}
\caption[]{${\cal R}_{L\,L^\prime}(T)$ of equation~\protect{(\ref{eq:8})}
against $T$, for $L^\prime=L+2$. Top to bottom: $L=4, 6, \dots 12$ 
}
\label{fig:5}
\end{center}
\end{figure}

For case (a), the low-temperature behaviour is similar to that found
in Section~\ref{sec:smg}, in that all curves (i) very nearly collapse on 
top of each other, 
and (ii) behave as $\exp(-1/T)$. Analogously to the ratio of scaled 
mass gaps ${\cal Q}_{L\,L^\prime}$ used in that  Section, we define
the  ratio of scaled free energies:
\begin{equation}
{\cal R}_{L\,L^\prime}(T) \equiv~\left[\,L\,\Delta f_L(T)\,\right]/
\left[\,L^\prime\,\Delta f_{L^\prime}(T)\,\right]\ ,\qquad {\rm (F)}  
\label{eq:8}
\end{equation}
Results, using $L^\prime=L+2$, are displayed in Figure~\ref{fig:5} 
and are very similar to those shown in  Figure~\ref{fig:2}, except that
instead of crossing the ${\cal R}=1$ axis from below, the curves only get 
asymptotically close to it, and from above. This is expected, as
inverse correlation length and domain-wall free energy are dual to each 
other~\cite{cardy84}. The values of
${\cal R}_{L\,L^\prime}$ at which the curves stabilize are, 
respectively, $1.03057$, $1.01029$, $1.00470$, $1.00253$, $1.00152$, 
for  $L=4, 6, \dots 12$. This sequence behaves as ${\cal 
R}_{L\,L^\prime}=e+f\,L^{-z}$,
where $z \simeq 2.7$ and $e$ equals unity within two parts in $10^4$.
Though the absence of a crossing of the ${\cal R}=1$ axis deprives one of 
an obvious reference point, it is possible to infer the limiting 
extent of the flat region, as $L \to \infty$, as follows. For each curve
one records at what temperature it reaches within, say, $10^{-4}$ of its
own $T \to 0$ asymptotic value, and the resulting sequence is 
extrapolated against $1/L$. By doing so we get $T_c=0.30 \pm 0.04$
where the error bar, though somewhat subjective, is certainly generous
enough to allow for the degree of arbitrariness involved in this 
procedure.  

We conclude that the evidence provided is entirely consistent with
that given by the analysis of correlation lengths in the F direction.
As a final remark, recall that with the TM in the F direction, the seam
whose energy is calculated lies parallel to ferromagnetic chains. With
$q=3$, the substitution in equation~(\ref{eq:7}) allows for a zero-energy
seam at $T=0$, hence the behaviour depicted in Fig.~\ref{fig:4} (a).  

As regards data displayed in Figure~\ref{fig:4} (b),
one finds well-defined fixed points of equation~(\ref{eq:6}). With 
$L^\prime=
L+2$, these are located respectively at $T^\ast=0.58753$, $0.52589$,
$0.48334$, $0.45108$, $0.42543$, for $L=4, 6, \dots 12$. When plotted
against $L^{-1}$, this sequence displays accentuated downward curvature.
Attempts to fit it against $L^{-t}$ produce $t \simeq 0.1$, with a {\em 
negative}
extrapolated value for $T_c$. A two-point fit of $L=10$ and $12$ data
against $L^{-1}$ gives an estimate $T_c=0.29$. Together with
the downward curvature of the full set of data, this indicates 
that the actual $T_c$ given by this particular domain-wall scaling is {\em 
below} $0.29$. 

We recall that in the previous case with the TM along the 
F direction, the corresponding domain walls were found to be critical for 
all $0 \leq T \leq T_c  \simeq 0.30$. 
With the TM along the AF direction the seam runs perpendicular to that 
least energetic orientation of domains, breaking ferromagnetic bonds as it 
goes. It would  thus appear that, if there is a well-defined temperature 
for the softening of this latter type of domain walls, it should be 
finite. However, we have not managed to produce a reliable estimate
for this from our sequence of fixed points. Bearing in mind
that the only typical non-zero temperature arising so far in the problem
is $T_c  \simeq 0.30$, one might interpret the data as signalling that
the extrapolated critical point for domain walls along the AF direction is 
{\em exactly} $T=0$. This brings new problems, as consideration
of the physical domain-wall free energy {\em per site}, $\Delta F_L = 
T\,\Delta  f_L/L$ shows that it behaves as $1-a(L)\,T$ for $T \to 0$, with 
$a(L) \simeq 2.7\,\exp (-4.6/L)$. In other words, at low temperatures 
it becomes extremely expensive to excite a domain wall along this direction. 
While this behaviour is, in a loose sense, dual to that of the correlation 
length along the AF direction, which remains finite at $T=0$
as given by equation~(\ref{eq:4}), it is hard to reconcile with the
idea that the respective domain walls would go soft at $T=0$. At present
we are not able to offer a consistent interpretation for the behaviour
just exhibited.     
 
\section{Approach to $T_c$ from above: finite-size scaling} 
\label{sec:tc}
Next we attempt a characterization of the transition from the paramagnetic
to the critical phase. It is usually assumed that, as this point is
approached from above in a Kosterlitz-Thouless transition, the bulk 
correlation length behaves as~\cite{gb92}:
\begin{equation}
\xi (T) =a_\xi\, \exp \left[ b_\xi\,(T-T_c)^{-\sigma}\right]\ ,
\label{eq:9}
\end{equation}
with $\sigma=1/2$, and $a_\xi$, $b_\xi$ and $T_c$ are non-universal.
From finite-size scaling~\cite{barber}, the correlation
length on a strip is expected to scale as:
\begin{equation} 
L\,\xi_L^{-1}(T) = f \left( L\,\exp 
\left[-b_\xi\,(T-T_c)^{-\sigma}\right]\right)\ .
\label{eq:10}  
\end{equation}
We have taken our data for correlation lengths calculated with the TM 
along the F direction, displayed in 
Fig.~\ref{fig:1} (a) as $L\,\xi_L^{-1}(T)$, 
and attempted to fit them to the form given by
equation~(\ref{eq:10}). We used the procedure described in~\cite{bs01},
to try and collapse our results for assorted values of $L$ and $T$ onto
a single curve. We took as  basic sets of data the six sequences taken 
each for fixed $L=4, 6, \dots 14$  against varying temperature, from 
$T=0.04$ to $1.1$. According to~\cite{bs01}, we then  took each of these 
sets in turn as a baseline, against which to fit the five other sets.
Since equation~(\ref{eq:10}) is valid only above $T_c$, the actual
number of data used would vary, for each trial fit, depending on the
particular value taken for the critical temperature. With $T_c$ in the
range $0.30-0.50$, as explained below,  the number of points fitted
was between $280$ ($T_c \simeq 0.30$) and $230$ ($T_c \simeq 0.50$).
For the local fitting procedure~\cite{bs01}, we used cubic
polynomials. 
 
We started by keeping $\sigma=1/2$ fixed, and varying $b_\xi$ and $T_c$.
From Sections~\ref{sec:smg} and~\ref{sec:dws}  
$T_c=0.30$  would seem a reasonable initial guess. Inspired in work done 
on  the $XY$ model~\cite{gb92},  $b_\xi$ was initially assumed of order 
unity. The landscape we found in parameter space, for the sum of 
residuals~\cite{bs01} as a function of  $b_\xi$ and $T_c$, was similar
to that described in~\cite{gb92}, with ``a number of local minima ...
at which ... minimization programs can get stuck''. Furthermore, visual
inspection of plots produced with the parameters set at the minimizing
values gave consistently poor fits in all cases for which $T_c$ 
was below $0.35$. We did, however, manage to find an overall minimum
for $T_c=0.51 \pm 0.01$, $b_\xi = 1.72 \pm 0.01$ (with none for 
$0.35 \leq T_c \leq 0.50$). 

\begin{figure}
\epsfxsize=8.5cm
\begin{center}
\epsffile{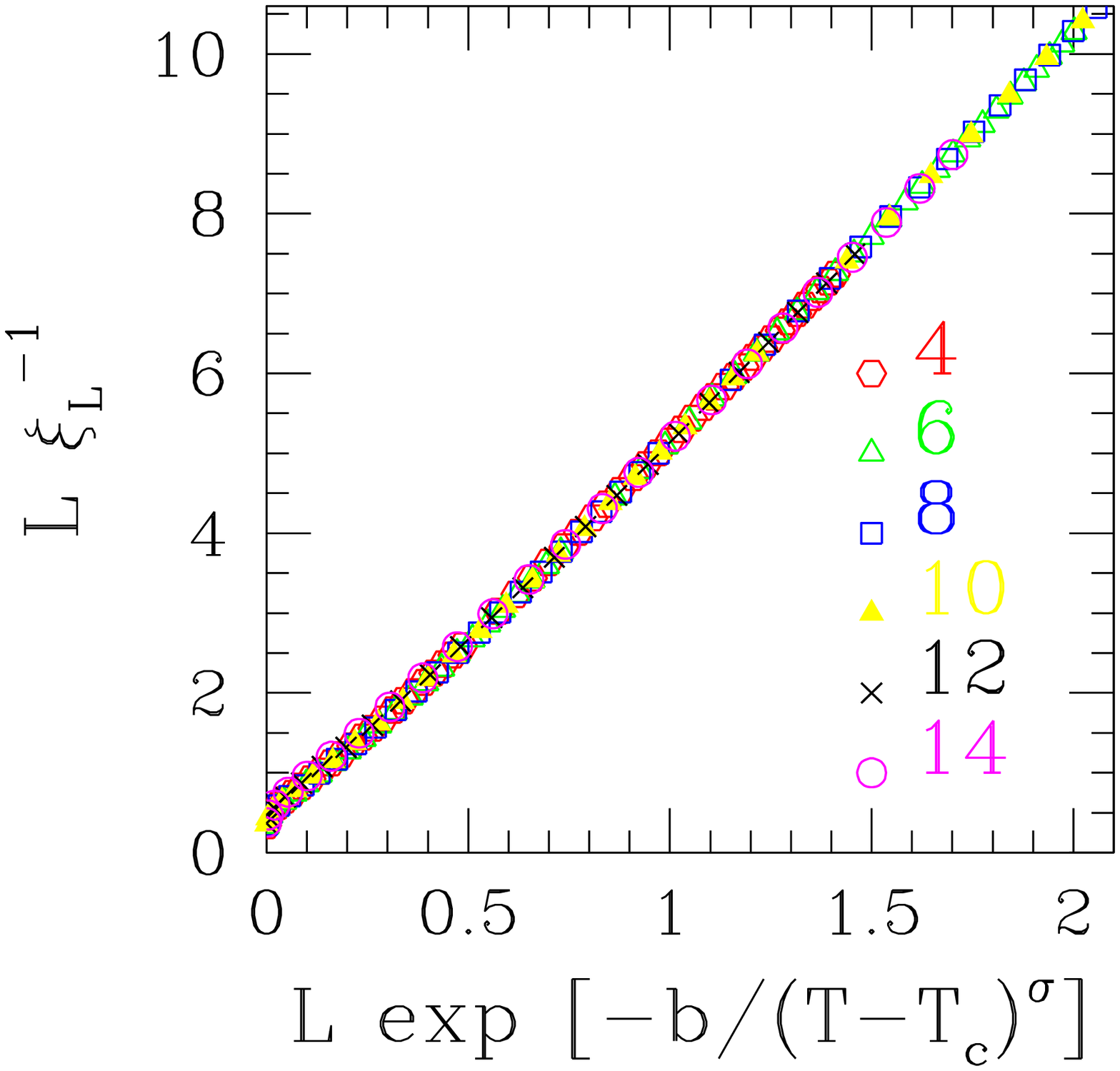}
\caption[]{Scaling plot (see equation~(\protect{\ref{eq:10}})) of data
displayed in Fig.\protect{\ref{fig:1}} (a). The parameters are:
$T_c=0.51$, $b_\xi=1.60$, $\sigma=0.54$ (see text). 
}
\label{fig:6}
\end{center}
\end{figure}

We then allowed $\sigma$ to vary. The overall picture was entirely
similar to that just described for fixed $\sigma=1/2$. The only
solution which gave a visually confirmed good fit was $T_c=0.50 \pm 0.01$, 
$b_\xi = 1.60 \pm 0.01$, $\sigma \simeq 0.54$. The corresponding
results are shown in Fig.~\ref{fig:6}.
The value $T_c \simeq 0.50$ is inconsistent with the evidence exhibited
in Sections~\ref{sec:smg} and~\ref{sec:dws}. One possible solution
to this contradiction would be if the critical region, where the
behaviour of equation~(\ref{eq:10}) is valid, 
were actually very narrow. Thus,
we would be distorting the picture, by using high-temperature 
data which in fact are outside the critical region. For the $XY$ model
in two dimensions, it has been argued that this is indeed the 
case~\cite{gb92,ggsm,car82}, with the width of the critical region
estimated as $(T-T_c)/T_c \leq 10^{-2}$~\cite{car82}. With $T_c \simeq
0.3$, in the present case this would mean a temperature interval $\Delta T 
\simeq 3 \times 10^{-3}$. 
For the fitting shown in Fig.~\ref{fig:6} we used data for 
$T$ as high as $2.5$, for the narrowest strips. We noticed that,
if only a subset of data with $T \leq 1.1$ is used, the best fit is found
for $T_c \simeq 0.47$, and is of similar quality to that of 
Fig.~\ref{fig:6}. Thus a plausible scenario is one in which, as only data 
for an ever narrowing interval above $T_c$ are considered, the 
corresponding fits give rise to lower and lower estimates of $T_c$ itself.
Should the prediction of~\cite{car82} hold true in the present case,
one would have a problem in that the accuracy to which $T_c$ is known so 
far (taking, say, our best estimate $0.29 \pm 0.01$ of 
Section~\ref{sec:smg}) is lower than the width $\Delta T$ of the critical
region.   

\section{Conclusions} 
\label{sec:conc}
We have provided strong numerical evidence that a critical
(Kosterlitz-Thouless) phase is present in the three-state mixed Potts
magnet described by equation~(\ref{eq:1}). Our argument relies on the
analysis of the quantities ${\cal Q}_{L\,L^\prime}(T)$ and ${\cal
R}_{L\,L^\prime}(T)$, defined respectively in equations~(\ref{eq:3})
and~(\ref{eq:8}).  Based on data exhibited in Figures~\ref{fig:2}
and~\ref{fig:5} and their respective extrapolation, we conclude that the
upper limit of the critical phase, above which the system behaves
paramagnetically, is $T_c=0.29 \pm 0.01$ (from Figure~\ref{fig:2};
treatment of data in Figure~\ref{fig:5} produces the less accurate, but
not inconsistent, estimate $T_c=0.30 \pm 0.04$).

Analysis of the (extremely anisotropic) nature of excitations below $T_c$,
as depicted in Figure~\ref{fig:3}, gives credence to the idea that the
critical phase extends all the way down to $T=0$, similarly to what
happens in the two-dimensional $XY$ model.

We have found that, while domain walls parallel to the ferromagnetic
direction are soft for the whole extent of the critical phase, those along
the antiferromagnetic direction seem to undergo a softening transition at
a well-defined, finite temperature. However, from our data we have not
been able to produce a reliable estimate for location of such transition.

Attempts to fit our data for $T > T_c$ to the finite-size scaling form of 
equation~(\ref{eq:10}) have met with qualified success. While we
have been able to produce rather good collapse plots, as depicted in
Figure~\ref{fig:6}, this has been done at the price of setting
$T_c=0.50 \pm 0.01$ which is clearly inconsistent with our own
earlier estimates. We have proposed a scenario in which such result
is interpreted as a distortion due to the narrowness of the critical 
region, similarly to the case of the $XY$ model~\cite{gb92,ggsm,car82}.
This is qualitatively borne out by the fact that fits with only
a subset of data, taken at not too high temperatures, indeed result
in a lower estimate for $T_c$. 

\ack
MQ thanks PIBIC/CNPq for the award of an undergraduate research 
fellowship; research of SLAdQ is partially supported by the 
Brazilian agencies CNPq (Grant No. 30.1692/81.5), FAPERJ (Grants
Nos. E26--171.447/97 and E26--151.869/2000) and FUJB-UFRJ.


\end{document}